\documentclass[aps,groupedaddress,10pt,pra,showpacs]{revtex4-2}
\usepackage{amsmath}
\usepackage{amssymb}
\usepackage{graphicx}
\usepackage{epstopdf}
\usepackage{color}
\usepackage{slashed}
\begin{document}

\preprint{Unknown}

\title{On heat coefficients, multiplicative anomaly and 4D Casimir energy for GJMS operators}
\author{Rodrigo Aros}
\email{raros@unab.cl}
\affiliation{Departamento de Fisica y Astronomia, Universidad Andres Bello,
Sazie 2212, Piso 7, Santiago, Chile}
\author{Fabrizzio Bugini}
 \email{fbugini@ucsc.cl}
\affiliation{Departamento de Matem\'atica y F\'isica Aplicadas,
Universidad Cat\'olica de la Sant\'isima Concepci\'on, Alonso de Ribera 2850, Concepci\'on, Chile}
\author{Danilo E. Diaz}%
\email{danilodiaz@unab.cl}
\affiliation{Departamento de Fisica y Astronomia, Universidad Andres Bello, Autopista Concepcion-Talcahuano 7100, Talcahuano, Chile}
\author{Camilo N\'u\~nez-Barra}
\email{cnb@uc.cl}
\affiliation{Facultad de Física, Pontificia Universidad Católica de Chile, Avenida Vicuña Mackenna 4860, Santiago, Chile}
\date{\today}

\begin{abstract}
This note aims to verify a prediction on the total derivative term of the 4D trace anomaly, and the corresponding heat coefficient, for GJMS operators. It stems from the explicit computation of an {\it improved} Casimir (or vacuum) energy on the sphere that takes into account the multiplicative anomaly among the (shifted) Laplacian factors and connects, via the Cappelli-Coste relation, with both the type A central charge and the total derivative term of the 4D trace anomaly. The present heat coefficient computation is based on Juhl's explicit formula for GJMS operators, Gilkey's formula for the integrated heat coefficient of higher-order Laplacians, and the \textit{conformal principle} by Branson and {\O}rsted.       
\end{abstract}

\maketitle

\section{Introduction}\label{sec:level1}

GJMS operators~\cite{GJMS92} are conformally covariant powers of the Laplacian originally envisaged within the Fefferman-Graham ambient construction~\cite{FG1984,Fefferman:2007rka} in conformal geometry. Together with the closely related notion of Q-curvature~\cite{Branson1993}, they have been the subject of much research over the last decades. The advent of the AdS/CFT correspondence in physics~\cite{Maldacena:1997re,Gubser:1998bc,Witten:1998qj} renewed interest in the Fefferman-Graham construction and has led, among many other interesting findings, to \textit{holographic} descriptions of both Q-curvature and GJMS operators in the associated Poincaré-Einstein metrics. Early developments related GJMS operators with scattering poles~\cite{GrahamZworski:2001} and allowed to express Q-curvatures explicitly in terms of volume coefficients~\cite{GrahamJuhl:2007}. Despite the intrinsic complexity of these constructs, these initial \textit{holographic} insights culminated in notable explicit and recursive formulae for GJMS operators and Q-curvatures due to Juhl \cite{Juhl_2013} (see also Fefferman and Graham \cite{FeffermanGrahamOnJuhl2012}).

Since functional determinants of GJMS operators naturally arise at one-loop quantum level in AdS/CFT correspondence, their study opens a window into the program of \textit{precision holography}. The central charges or trace anomaly coefficients~\cite{Capper:1974ic} encoded in the heat-kernel coefficients are primordial. The latter poses a challenge to traditional heat-kernel techniques due to their higher derivative nature; however, due to factorization properties on Einstein backgrounds~\cite{Gover:2006}, their so-called type A central charge has been derived in generic even dimensions~\cite{Diaz:2007an,Diaz:2008hy,Dowker:2010qy}, while their type B central charge~\cite{Deser1993GeometricCO} has been obtained in $4$ and $6$ dimensions\cite{Beccaria:2014xda,Beccaria:2017dmw,Bugini:2018def}.   
The accumulative features of the heat-kernel coefficients of the individual Laplacian factors is what allows to bypass detailed knowledge of the heat-kernel coefficient for the high-derivative GJMS operators. 
In the present work, we focus on the four-dimensional case, where the trace anomaly reads
\begin{eqnarray}
    {\cal A}\,&=&\,-a\,E_4\,+\,c\,W^2\,-\,g\,\Delta R\\\nonumber\\
&=&\,-4a\,Q_4\,+\,(c-a)\,W^2\,+\,\gamma\,\Delta J~.
\end{eqnarray}
In the second line we traded, as a preparation, the Euler density by the Q-curvature and the Ricci scalar by the Schouten one~\footnote{In our convention $\Delta=-\nabla^2$ denotes the non-negative Laplacian.}. It is convenient to explicitly display the available data on the {\it universal} (i.e. regularization-scheme independent) quantities 
\begin{equation}
    a\,=\,\frac{k^3}{144}-\frac{k^5}{240}\qquad,\qquad c-a\,=\,\frac{k}{180}.
\end{equation}
We concentrate our attention on the coefficient $\gamma$ of the total derivative term. It is well known that it can be modified by the addition of a finite counter term, namely its conformal primitive, which is the volume integral of $J^2$. Therefore, it is a regularization-scheme-dependent quantity. This ambiguity is inherited by the Casimir or vacuum energy on spheres that dominates the low temperature ($\beta\rightarrow\infty$) asymptotics of the partition function (functional determinant) on $S^1_\beta\times S^3$ via the Cappelli-Coste relation~\cite{CAPPELLI1989707}   
\begin{equation}
    {\cal E}_c\,=\,a\,-\,\frac{1}{16}\gamma~.
\end{equation}
The accumulated Casimir energy for GJMS, as computed via standard zeta function regularization, reads~\cite{Beccaria:2014xda} 
\begin{equation}
    {\cal E}_{c,acc}\,=\,-\dfrac{k}{720}\left(6k^4-20k^2+11\right)
\end{equation}
and grants access to the coefficient of the total derivative
\begin{equation}
    \gamma_{acc}\,=\,\dfrac{k}{45}\left(3k^4-15k^2+11\right)~.
\end{equation}
The value for the conformal Laplacian ($k=1$) agrees with the value resulting from the standard heat kernel computation of $b_{4,4}$. However, for the Paneitz operator ($k=2$) there is a discrepancy with the explicit heat kernel coefficient $b_{4,4}$ reported in the literature. The latter follows either from the explicit heat coefficient originally obtained by Gusynin~\cite{GUSYNIN1989233} for quartic operators, or from the Polyakov formula for the determinant of the Paneitz operator as derived by Branson~\cite{Branson:1996bp}. 
The discrepancy is removed, to our surprise, when we consider instead the \textit{improved} Casimir energy~\cite{Aros:2023hgi} that takes into account the multiplicative anomaly between the Laplacian factors  
\begin{equation}
    {\cal E}_{c}\,=\,-\dfrac{k^3}{720}\left(2k^2-5\right)
\end{equation}
leading to the following prediction for the total derivative coefficient
\begin{equation}
    \gamma\,=\,-\dfrac{k^5}{45}.
\end{equation}
Our present purpose is to complete the calculation of the heat coefficient $b_4$ by following the program initiated by Branson \cite{Branson:1996bp} for the Paneitz operator in 4D and extending it to the whole family of GJMS operators. We restrict to Bach-flat metrics to avoid obstructions and work out, based on Juhl's explicit formula, the necessary building blocks that enter the integrated heat coefficient as devised by Gilkey \cite{Gilkey} long ago. The conformal principle of Branson and {\O}rsted~ allows them to read off the coefficient of the total derivative $\Delta J$ from the necessarily vanishing coefficient of $J^2$, once the heat kernel is expressed in the Q-curvature basis, by continuation in the dimension. To our knowledge, there are no analog results like that of Gusynin for higher than quartic differential operators (see, however, Barvinsky et alia \cite{Barvinsky:2024kgt,Barvinsky:2021ijq} for recent progress).   

\section{Preliminaries}

\subsection{Gilkey's heat coefficient for higher order Laplacians}
Let us start by recalling Gilkey's \cite{Gilkey} major result on the heat coefficient for higher-order Laplacians, restricted to the scalar case for our present purposes.  
For a natural and homogeneous differential operator ${\cal P}$ of order $u=2\,v\geq4$, with leading symbol given by $v-$th power of the metric tensor,  of the form 
\begin{equation}
{\cal P}\,=\,(\Delta)^v\,+\,p_{2,ij}\nabla^ i\nabla^j(\Delta)^{v-2}\,+\,(-1)^v\left\{p_{3,i_1...i_{u-3}}\nabla_{i_1}...\nabla_{i_{u-3}}\,+...+\,p_u\right\}~,
\end{equation}
Gilkey~\cite{Gilkey} has shown that its -diagonal and integrated on a closed $n-$dimensional Riemannian manifold $\mathfrak{M}_n$- heat coefficient $B_4$ takes the form
\begin{eqnarray}
(4\pi)^{n/2}\,v\,B_{4,n}[\mathcal{P}] &=&\,\dfrac{\Gamma(\frac{n-4}{u})}{\Gamma(\frac{n-4}{2})}\cdot\frac{1}{360\,v^2\,n(n^2-4)}\int_{\mathfrak{M}_n}dvol_g\, v^2\,n(n^2-4) [2\,Riem^2\,-\,2\,Ric^2\,+\,5\,R^2 \nonumber \\ 
    &+& \,60\,v\,n(n+2)\,R\cdot p_2,_i^{\;i}\nonumber\\
    &-& 120\,v\,n(n+2)\,R^{ij}\cdot p_2,_{ij}\,+\,180\,(2v+n-4)(p_2,_i^{\;i})^2\,+\,360\,(2v+n-4) |p_2|^2\nonumber\\&-&\,720\,v\,n(n+2)\,S(p_4)/S(\delta^{v-2})~.
\end{eqnarray}
In the above, $S(...)$ stands for the symmetrized trace and our convention for Riemann and Ricci tensor and scalar differs by a sign.   
A few remarks are in order now. First, the coefficients depend explicitly and non-trivially on the dimension $n$ as opposed to the standard second-order case. Second, only $p_2$ and $p_4$ appear in the formula for the integrated heat coefficient, any further possible dependence on the rest of the $p's$ can only arise in pure divergence terms, \textit{i.e.}, total derivatives.     
In general, thus, one has no access to the total derivative terms of the heat coefficient. However, when conformal invariance comes into play then more can be said about the heat coefficients for higher-order conformal Laplacians and one can reconstruct the local information from the integrated information via a continuation in the dimension argument introduced by Branson and {\O}rsted. 

\subsection{Branson-{\O}rsted's conformal principle}
Consider now the short-time asymptotic expansion of the diagonal heat kernel 
\begin{equation}
    \mbox{Tr}\,\left(f\,e^{-t\,{\cal P}}\right)\sim \sum_{j=0}^{\infty} t^{(j-n)/u}\int_{\mathfrak{ M}_n}dvol_g\left(f\,b_{j,n}[{\cal P}]\right)~,
\end{equation}

If in addition $\cal P$ is conformally covariant, in the sense that under a local (Weyl) rescaling of the metric $g\rightarrow e^{2\,w}\,g$ it follows that ${\cal P}\rightarrow e^{-b\,w}{\cal P}[e^{a\,w}]$, or if $\cal P$ is a positive integral power of a conformal covariant, then the infinitesimal variation ($\epsilon\cdot w$) of the integrated heat coefficient satisfies~ 
\begin{equation}
\left.\frac{d}{d\epsilon} \right|_{\epsilon=0}\int_{\mathfrak{M}_n}\left(dvol_g\,b_{j,n}[{\cal P}]\right)_{\epsilon\cdot w} \,=\,(n-j)\cdot\int_{\mathfrak{M}_n}w\cdot\left(dvol_g\,b_{j,n}[{\cal P}]\right)_0~.
\end{equation}
It follows, in particular, the global conformal invariance of the integrated \textit{critical} heat coefficient $B_{n,n}[{\cal P}]=\int_{\mathfrak{M}_n}dvol_g\,b_{j,n}[{\cal P}]$. 

Branson~\cite{Branson:1996bp} has used this conformal principle combined with further insights from Weyl's invariant theory on the structure of the conformal invariants to reconstruct the total derivative term that is washed away by the volume integration. In a suitable basis for conformal invariants, where Euler density is traded off by the Q-curvature $Q_{4,n}$, the heat coefficient can be expanded as follows   
\begin{equation}
    b_{4,n}[{\cal P}]=\beta_0\cdot J^2\,+\,\beta_1\cdot W^2\,+\,\beta_2\cdot Q_{4,n}\,+\,\beta_3\cdot \Delta J~.
\end{equation}
Upon the infinitesimal variation, there is a remainder that must vanish under the conformal assumption and leads to the coefficient of the \textit{hidden} total derivative 
\begin{equation}
\left\{-2\beta_0+(n-4)\beta_3\right\}\cdot\int_{\mathfrak{M}_n}w\cdot\left(dvol_g\,\Delta J\right)_0
\end{equation}
The analytical dependence in the dimension $n$ is crucial. The vanishing of the $\beta_0$ coefficient for the critical $b_{4,4}$ comply with the Deser-Schwimmer assertion~\cite{Deser1993GeometricCO} (see also Branson et alia~\cite{Branson1995InvariantsOL}) for the trace anomaly.
The corresponding value of the total derivative coefficient for the conformal Laplacian or Yamabe operator can be readily verified by standard heat kernel results. For the Paneitz operator, in turn, the integrated critical heat coefficient can be obtained from Gilkey's formula displayed above and the coefficient of the total derivative obtained by Branson matches the explicit term for quartic operators derived by Gusynin~\cite{GUSYNIN1989233}. To our knowledge, there are no explicit results for the total derivative term for operators of higher order, and the best information is still provided by Gilkey's seminal work that washes away the total derivative term. 

\subsection{Juhl's explicit formula for GJMS operators}
To complete Branson's program, we will make use of the following explicit formula for GJMS operators as a linear combination of compositions of natural second-order differential operators derived by Juhl~\cite{Juhl_2013}
\begin{equation}
    {\cal P}_{2k}=\sum_{|I|=N}n_I\cdot{\cal M}_{2I}
\end{equation}
valid for $k\geq1$ (and $k\leq n/2$ if n is even)~\footnote{Note that our convention differs from that of Juhl's, our $\Delta=-\nabla^2$ is the non-negative Laplacian and in consequence, our ${\cal P}_{2k}$ will differ by a sign for odd $k$. Our convention seems better suited to Gilkey's formula.}. The multi-index $I$ runs from $(1,...,1)$, accompanying the composition of the building blocks ${\cal M}_2\circ...\circ{\cal M}_2$, to $(N)$ with ${\cal M}_{2N}$. For a given sequence $I=(I_1,...,I_r)$ of natural numbers, the integer coefficient $n_I$ is given by 
\begin{equation}
    n_I=\prod_{j=1}^r\binom{\sum_{k\leq j}I_k-1}{I_j-1}\cdot\binom{\sum_{k\geq j}I_k-1}{I_j-1}
\end{equation}
Below we list the few building blocks relevant to our present analysis
\begin{eqnarray}
    {\cal M}_2&&=\Delta\,+\,\mu_2\quad\,,\,\quad\mu_2=\frac{n-2}{2}J\\
    \nonumber\\
    {\cal M}_4&&=4\,\nabla^i\,P_{ij}\,\nabla^j\,+\,\mu_4\quad\,,\,\quad\mu_4=-J^2-(n-4)|P|^2-\Delta J\\
    \nonumber\\
    {\cal M}_6&&=-48\,\nabla^i\,P_{im}P_{j}^{\,\,m}\,\nabla^j\,-\dfrac{16}{n-4}\,\nabla^i\,B_{ij}\,\nabla^j\,+\mu_6
\end{eqnarray}
In the following, we restrict to Bach-flat metrics so that the Bach tensor $B$ above vanishes identically. Moreover, we do not need the explicit expression for $\mu_6$, it is enough to notice that for dimensional reasons it cannot enter the formula for the $B_{4,n}$ heat coefficient. We do not need either to keep track of the $\nabla^2J$ of $\mu_4$, it will be washed away by the  
volume integral in Gilkey´s formula. 

\section{Heat coefficient $b_{4,n}$ : take I}
Let us start by examining the cases where the explicit formula lends itself to explicit calculations before we address the general case.  

\subsubsection{Conformal Laplacian or Yamabe operator: $P_2$}
We begin with 
\begin{equation}
{\cal P}_2={\cal M}_2=\Delta\,+\,\mu_2=\Delta\,+\,\frac{n-2}{2}J
\end{equation}

\begin{equation}
    (4\pi)^{n/2}\,B_{4,n}[{\cal P}_2]\,=\,\int_{\mathfrak{M}_n}dvol_g\,\left\{\dfrac{1}{360}\left[2\,Riem^2\,-\,2\,Ric^2\,+\,5\,R^2\right] \,-\,\frac{1}{6}\,R\,\mu_2\,+\,\frac{1}{2}\,(\mu_2)^2\right\}~.
\end{equation}
Now we go to the Q-curvature basis using the following identities
\begin{eqnarray}
&&R=2(n-1)J\qquad,\qquad R^2=4(n-1)^2J^2\qquad,\qquad Ric^2=(n-2)^2|P|^2+(3n-4)J^2\qquad,\\\nonumber\\
&&Riem^2=W^2+4(n-2)|P|^2+4J^2\qquad\mbox{and}\qquad Q_{4,n}=\frac{n}{2}J^2-2|P|^2+\Delta J~,
\end{eqnarray}
and obtain
\begin{equation}
    (4\pi)^{n/2}\,B_{4,n}[{\cal P}_2]\,=\,\int_{\mathfrak{M}_n}dvol_g\,\left\{-\frac{(n-4)(n-6)(n-8)}{720}\cdot J^2\,+\,\frac{1}{180}\cdot W^2\,+\,\dfrac{(n-2)(n-6)}{360}\cdot Q_{4,n}\right\}~,
\end{equation}
and finally, by the conformal principle, we complete the local information on the heat coefficient
\begin{equation}
    (4\pi)^{n/2}\,b_{4,n}[{\cal P}_2]\,=\,\dfrac{(n-2)(n-6)}{360}\cdot Q_{4,n}\,+\,\frac{1}{180}\cdot W^2\,-\,\frac{(n-6)(n-8)}{360}\cdot\left[\Delta J+\frac{n-4}{2}\,J^2\right]~.
\end{equation}
In particular, for the 4D trace anomaly 
\begin{equation}
   (4\pi)^2 {\cal A}_4=-4a\cdot Q_{4}\,+\,(c-a)\cdot W^2\,+\,\gamma\cdot\Delta J
\end{equation}
we confirm the standard result
\begin{equation}
   (a, c-a,\gamma)= (\frac{1}{360},\frac{1}{180},-\frac{1}{45})~.
\end{equation}
This first example illustrates the consistency of the conformal principle. The $\beta_3\cdot\Delta J$ term for the conformal Laplacian is well known to be given by $-\frac{1}{30}\Delta R+\frac{1}{6}\Delta \mu_2=\frac{n-6}{60}\Delta J$ but we need to include the additional contribution from the total derivative of the Q-curvature $-\beta_2\cdot\Delta J=-\frac{(n-2)(n-6)}{360}\cdot\Delta J$ after the change of basis, the overall result correctly matches the $-\frac{(n-6)(n-8)}{360}\cdot\Delta J$ of the heat coefficient and the $-\frac{1}{45}\cdot\Delta J$ of the trace anomaly as previously obtained.
In addition, we also verify our expectation for $\gamma$ from the Cappelli-Coste~\cite{CAPPELLI1989707} relation and the Casimir energy, namely $\left.-\frac{k^5}{45}\right|_{k=1}=-\frac{1}{45}$. 

\subsubsection{Conformal squared-Laplacian or Paneitz operator: $P_4$}
We now follow the lead of Branson in the case of the quartic operator. We find it instructive to start with Juhl's explicit formula

\begin{eqnarray}
{\cal P}_4\,=\,{\cal M}_2^2\,+\,{\cal M}_4&&\,=\,\left\{\Delta^2\,+\,2\mu_2\Delta-2(\nabla^i\mu_2)\nabla_i+\mu_2^2+(\Delta \mu_2)\right\}+\left\{4P^{ij}\nabla_i\nabla_j +4(\nabla^iJ)\nabla_i+\mu_4\right\}\\\nonumber\\\nonumber
&&\,=\,\Delta^2+\left[4P^{ij}-2\mu_2g^{ij}\right]\nabla_i\nabla_j+\left[\mu_2^2+\mu_4\right]+...~.
\end{eqnarray}
In the last line above we have only kept the terms relevant to Gilkey's formula, which now reads
\begin{eqnarray}
    (4\pi)^{n/2}\,2\,B_{4,n}[{\mathcal{ P}}
_4]\,&=&\,\dfrac{\Gamma(\frac{n-4}{4})}{\Gamma(\frac{n-4}{2})}\cdot\frac{1}{360\,(n^2-4)}\int_{\mathfrak{M}_n}dvol_g\,\left\{(n^2-4)\left[2\,Riem^2\,-\,2\,Ric^2\,+\,5\,R^2\right] \,+\,30\,(n+2)\,R\cdot p_2,_i^{\;i}\right.\nonumber\\\nonumber\\
    &&\left.\,-\,60\,(n+2)\,R^{ij}\cdot p_2,_{ij}\,+\,45\,(p_2,_i^{\;i})^2\,+\,90\,|p_2|^2\,-\,360\,(n+2)\,p_4\right\}~.
\end{eqnarray}
From the explicit formula for ${\cal P}_4$ we then read off
\begin{eqnarray}
p_{2,ij}\,&=&\,4\,P_{ij}-(n-2)\,J\,g_{ij}~,\\\nonumber\\
p_4&=&\frac{n-4}{2}\left(\frac{n}{2}\,J^2-2\,|P|^2\right)~,
\end{eqnarray}
where we have dropped a total derivative in $p_4$ that integrates to zero in Gilkey's formula. The necessary contractions go as follows
\begin{eqnarray}
p_2,_i^{\;i}\,&&=\,-(n^2-2n-4)\,J\quad,\quad(p_2,_i^{\;i})^2\,=\,(n^2-2n-4)^2\,J^2\quad,\quad|p_2|^2\,=\,16\,|P|^2+(n-4)(n^2-4)\,J^2\quad,\\\nonumber\\
&&R\cdot p_2,_i^{\;i}\,=\,-2(n-1)(n^2-2n-4)\,J^2\quad,\quad R^{ij}\cdot p_{2,ij}\,=\,4(n-2)\,|P|^2-2n(n-3)\,J^2~.
\end{eqnarray}
As in the case of the conformal Laplacian, we go to the Q-curvature basis and obtain
\begin{eqnarray}
   (4\pi)^{n/2}\,2\,B_{4,n}[{\mathcal{ P}}_4]\,&=&\,\dfrac{\Gamma(\frac{n-4}{4})}{\Gamma(\frac{n-4}{2})}\int_{\mathfrak{M}_n}dvol_g\,\left\{-\frac{(n-4)(n-8)(n-12)(n+4)}{720\,(n-2)}\cdot J^2\,+\,\frac{1}{180}\cdot W^2\right.\nonumber\\ \nonumber \\ 
   &+& \left.\,\dfrac{(n-8)(n^3-52n-24)}{360\,(n^2-4)}\cdot Q_{4,n}\right\},
\end{eqnarray}
and finally, by the conformal principle, we complete the local information on the heat coefficient
\begin{eqnarray}
(4\pi)^{n/2}\,2\,b_{4,n}[{\mathcal{P}}_4]\,&=&\,\,\dfrac{\Gamma(\frac{n-4}{4})}{\Gamma(\frac{n-4}{2})}
    \left\{\dfrac{(n-8)(n^3-52n-24)}{360\,(n^2-4)}\cdot Q_{4,n}\,+\,\frac{1}{180}\cdot W^2\right.\nonumber\\
    &-&\left.\frac{(n-8)(n-12)(n+4)}{360\,(n-2)}\cdot\left[\Delta J+\frac{n-4}{2}\,J^2\right]\right\}~.
\end{eqnarray}

In particular, for the 4D trace anomaly, we obtain 
\begin{equation}
   (a, c-a,\gamma)= \left(-\frac{7}{90},\frac{1}{90},-\frac{32}{45}\right)~.
\end{equation}
The above result can be independently verified using the formula by Gusynin~\cite{GUSYNIN1989233} for quartic operators that include the total derivative term. It can be traced back to the total derivative coming from the additional terms 
\begin{equation}
    \frac{180}{n-2}\left(-\frac{n-2}{15}\Delta R\,-\,\frac{n+4}{6(n+2)}\Delta p_2,_i^{\;i}-\frac{2}{3}\frac{n+1}{n+2}\nabla^i\nabla^jp_{2,ij}+\nabla^ip_3,i-2p_4\right)~,
\end{equation}
so that we need the complete structure of the Paneitz operator including the total derivative in the constant term $p_4$. These are given by 
\begin{eqnarray}
    \nabla^i\nabla^jp_{2,ij}\,&=&\,\nabla^i\nabla^j\left[4\,P_{ij}-(n-2)J\,g_{ij}\right]=(n-6)\Delta J~,\\\nonumber\\
    \nabla^ip_{3,i}\,&=&\,\nabla^i\left[(-n+6)\nabla_i J\right]=(n-6)\Delta J~,\\\nonumber\\
    p_4\,&=&\frac{n-4}{2}\left(\frac{n}{2}\,J^2-2\,|P|^2+\Delta J\right)~.
\end{eqnarray}
Finally, we need to include the additional contribution from the total derivative of the Q-curvature $-\frac{(n-8)(n^2-52n-24)}{360\,(n^2-4)}\cdot\Delta J$ after the change of basis. The overall result correctly matches the total derivative of the heat coefficient obtained by Branson and the $-\frac{32}{45}\cdot\Delta J$ of the trace anomaly.
Again, we also verify our expectation for $\gamma$ from the Cappelli-Coste relation \cite{CAPPELLI1989707} and the \textit{improved} Casimir energy~\cite{Aros:2023hgi}, namely $\left.-\frac{k^5}{45}\right|_{k=2}=-\frac{32}{45}$.

\subsubsection{Conformal cubed-Laplacian or Branson operator: ${\cal P}_6$}

To proceed further we examine Juhl's explicit formula \cite{Juhl_2013,FeffermanGrahamOnJuhl2012} for the conformal cubed-Laplacian, sometimes also called Branson operator. It is worth noticing and although there are several expressions in the literature for ${\cal P}_6$, that differ by conventions and also ambiguities, here we can safely restrict to conformally flat metrics where the obstruction and the ambiguities vanish \cite{Branson1985,Osborn:2015rna,paci2024}. 

\begin{eqnarray}
{\cal P}_6\,&=&\,{\cal M}_2^3\,+\,2\left\{{\cal M}_2{\cal M}_4\,+\,{\cal M}_4{\cal M}_2\right\}\,+\,{\cal M}_6 \\ \nonumber\\\nonumber
&&\,=\,\left\{\Delta^3\,+\,3\mu_2\Delta^2\,+\,3\mu_2^2\Delta\right\}\,+\,\left\{16P^{ij}\nabla_i\nabla_j\Delta\,+\,16\mu_2P^{ij}\nabla_i\nabla_j\,+\,4\mu_4\Delta \right\}+\,\left\{-48P^{i}_mP^{jm}\nabla_i\nabla_j\right\}+...~.
\end{eqnarray}
In the second line above we have again only kept the terms relevant for Gilkey's formula. It reads
\begin{eqnarray}
    (4\pi)^{n/2}\,3\,B_{4,n}[{\mathcal{ P}}_6] =\dfrac{\Gamma(\frac{n-4}{6})}{\Gamma(\frac{n-4}{2})}\cdot\frac{1}{360\,n(n^2-4)}\int_{\mathfrak{M}_n} && dvol_g \{n(n^2-4)\left[2\,Riem^2\,-\,2\,Ric^2\,+\,5\,R^2\right] \nonumber\\
    &+& 20\,n(n+2)\,R\cdot p_2,_i^{\;i} \nonumber\\
    &-& 40\,n(n+2)\,R^{ij}\cdot p_2,_{ij}\,+\,20\,(n+2)\,(p_2,_i^{\;i})^2\,+\,40\,(n+2)\,|p_2|^2\nonumber\\
    &-& 240\,(n+2)\,p_4,_i^{\;i} \}~.
\end{eqnarray}
From the excerpts of the explicit formula for ${\cal P}_6$ displayed above, we read off
\begin{eqnarray}
p_{2,ij}\,&=&\,16\,P_{ij}-\dfrac{3}{2}(n-2)Jg_{ij}~,\\\nonumber\\
p_{4,ij}\,&=&\,48P_{i}^mP_{jm}\,-\,8(n-2)JP_{ij}\,+\,\left[\frac{3n^2-12n-4}{4}J^2-4(n-4)|P|^2\right]g_{ij} ~,
\end{eqnarray}
where we have dropped terms in $p_{4,ij}$ that upon trace integrate to zero in Gilkey's formula. The necessary contractions go as follows
\begin{eqnarray}
p_2,_i^{\;i} &=& \,-\frac{3n^2-6n-32}{2}\,J \nonumber\\ 
(p_2,_i^{\;i})^2 &=& \frac{(3n^2-6n-32)^2}{4}\,J^2,\nonumber \\
|p_2|^2\,&=&\,256\,|P|^2+\frac{3}{4}(n-2)(3n^2-6n-64)\,J^2 \nonumber\\
\qquad R\cdot p_2,_i^{\;i}\,&=& -(n-1)(3n^2-6n-32)\,J^2 \nonumber\\
R^{ij}\cdot p_{2,ij}\,&=&\,16(n-2)\,|P|^2-(3n^2-9n-10)\,J^2 \nonumber\\
p_4,_i^{\;i}\,&=&\,-4(n+2)(n-6)|P|^2+\frac{3n^3-12n^2-36n+64}{4}\,J^2
\end{eqnarray}
Changing to the Q-curvature basis, we obtain
\begin{widetext}
\begin{eqnarray}
     (4\pi)^{n/2}\,3\,B_{4,n}[{\cal P}_6] &=& \,\dfrac{\Gamma(\frac{n-4}{6})}{\Gamma(\frac{n-4}{2})}\int_{\mathfrak{M}_n}dvol_g\,\left\{-\frac{(n-4)(n-10)(n-16)(n+2)(n+8)}{720\,n(n-2)}\cdot J^2 \right. \nonumber\\
     && \nonumber\\
     &+& \left. \frac{1}{180}\cdot W^2\,+\,\dfrac{(n-10)(n^3-132n-64)}{360\,n(n-2)}\cdot Q_{4,n}\right\}
\end{eqnarray}
\end{widetext}
and finally, by the conformal principle, we can complete the local information on the heat coefficient
\begin{widetext}
\begin{eqnarray}
        (4\pi)^{n/2}\,3\,b_{4,n}[{\cal P}_6] &=& \dfrac{\Gamma(\frac{n-4}{6})}{\Gamma(\frac{n-4}{2})}
    \left\{\dfrac{(n-10)(n^3-132n-64)}{360\,n(n-2)}\cdot Q_{4,n}\,+\,\frac{1}{180}\cdot W^2\right.\nonumber \\
    && \nonumber\\
    &-& \left.\frac{(n-10)(n-16)(n+2)(n+8)}{360\,n(n-2)}\cdot\left[\Delta J+\frac{n-4}{2}\,J^2\right]\right\}.
\end{eqnarray}
\end{widetext}

In particular, for the 4D trace anomaly we obtain 
\begin{equation}
   (a, c-a,\gamma)= (-\frac{33}{40},\frac{1}{60},-\frac{27}{5})~.
\end{equation}
We can only verify, besides the central charges $a$ and $c-a$, our expectation for $\gamma$ from the Cappelli-Coste relation and the {\it improved} Casimir energy, namely $\left.-\frac{k^5}{45}\right|_{k=3}=-\frac{27}{5}$. Yet, this simple result combines several ingredients and gives confidence in the correctness of all steps involved.

\subsubsection{Conformal fourth power of the Laplacian: ${\cal P}_8$} 
Let us continue with the rare case of ${\cal P}_8$ before attempting a promising generalization. We resort again to Juhl's explicit formula
\begin{widetext}
\begin{eqnarray}
{\cal P}_8\,&=&\,{\cal M}_2^4\,+\,\left\{3{\cal M}_2^2{\cal M}_4\,+\,4{\cal M}_2{\cal M}_4{\cal M}_2\,+\,3{\cal M}_4{\cal M}_2^2\right\}\,+\,9{\cal M}_4^2\,+\,3\left\{{\cal M}_2{\cal M}_6\,+\,{\cal M}_6{\cal M}_2\right\}\,+\,{\cal M}_8
\nonumber \\
&=&\left\{\Delta^4+4\mu_2\Delta^3+6\mu_2^2\Delta^2\right\}+\left\{40P^{ij}\nabla_i\nabla_j\Delta^2+80\mu_2P^{ij}\nabla_i\nabla_j\Delta+10\mu_4\Delta^2\right\}\nonumber\\
&+& \left\{144P^{ij}P^{kl}\nabla_i\nabla_j\nabla_k\nabla_l\right\}-\left\{288P^{im}P^{j}_m\nabla_i\nabla_j\Delta\right\}+\ldots.\nonumber
\end{eqnarray}
\end{widetext}
It is remarkable that starting with ${\cal M}_8$, none of the higher orders have any bearing in Gilkey's formula for the integrated heat coefficient. In the second line above we have again only kept the exclusive terms that enter Gilkey's formula
\begin{eqnarray}
    (4\pi)^{n/2}\,4\,B_{4,n}[{\mathcal{P}}_8] = \dfrac{\Gamma(\frac{n-4}{8})}{\Gamma(\frac{n-4}{2})}\cdot\frac{1}{360\,n(n^2-4)}&& \int_{\mathfrak{M}_n}dvol_g \left\{n(n^2-4)[
    2\,Riem^2\,-\,2\,Ric^2\,+\,5\,R^2] \right.\nonumber \\
    && +15\,n(n+2)\,R\cdot p_2,_i^{\;i} \nonumber\\
    && -30\,n(n+2)\,R^{ij}\cdot p_2,_{ij}\,+\,45\,(n+4)\,(p_2,_i^{\;i})^2\nonumber\\
    && +\left. 90\,(n+4)\,|p_2|^2\,-\,180\,n(n+2)\,S(p_4)/S(\delta^2)\right\}.
\end{eqnarray}
From the excerpts of the explicit formula for ${\cal P}_8$ displayed above, we again read off
\begin{eqnarray}
p_{2,ij}\,&=&\,40\,P_{ij}\,-\,2(n-2)Jg_{ij}~,\\\nonumber\\
p_{4,ijkl}\,&=&\,144P_{ij}P_{kl}\,+\,\left[\,288P_{i}^mP_{jm}-40(n-4)JP_{ij}\,\right]g_{kl}\,+\,\left[\frac{3n^2-12n-8}{2}J^2-10(n-2)|P|^2\right]g_{ij}g_{kl} ~,
\end{eqnarray}
where we have written $p_{4,ijkl}$ up to terms that upon trace integrate to zero on a closed manifold. The necessary contractions go as follows
\begin{widetext}
\begin{eqnarray}
p_2,_i^{\;i}\,& &=\,-2(n^2-2n-20)J \nonumber \\ 
&& \nonumber\\
(p_2,_i^{\;i})^2\,&=&\,4(n^2-2n-20)^2\,J^2\quad,\quad|p_2|^2\,=\,1600\,|P|^2\,+\,4(n-2)(n^2-2n-40)\,J^2 \\ & & \nonumber\\
R\cdot p_2,_i^{\;i}\,&=&-4(n-1)(n^2-2n-20)\,J^2\nonumber\\
&& \nonumber\\
R^{ij}\cdot p_{2,ij}\,&=&\,40(n-2)\,|P|^2\,-\,4(n^2-3n-8)\,J^2 \nonumber\\
&& \nonumber\\
S(p_4)\,&=&\,p_{4,ijkl}\,\frac{g^{ij}g^{kl}+g^{ik}g^{jl}+g^{il}g^{jk}}{3}\nonumber\\
&=& -\frac{16}{3}(5n^3-10n^2-184n-432)|P|^2+\frac{1}{6}(3n^4-6n^3-112n^2-6n+608)J^2 \nonumber\\
&& \nonumber\\
S(\delta^2)\,&=&\,g_{ij}g_{kl}\,\frac{g^{ij}g^{kl}+g^{ik}g^{jl}+g^{il}g^{jk}}{3}\,=\,\frac{n(n+2)}{3}~.
\end{eqnarray}
\end{widetext}
Changing to the Q-curvature basis, we obtain
\begin{widetext}
\begin{eqnarray}
    (4\pi)^{n/2}\,4\,B_{4,n}[{\cal P}_8]\,=\,\dfrac{\Gamma(\frac{n-4}{8})}{\Gamma(\frac{n-4}{2})}\int_{\mathfrak{M}_n}dvol_g\,&&\left\{-\frac{(n-4)(n-12)(n-20)(n+4)(n+12)}{720\,n(n-2)}\cdot J^2\,+\,\frac{1}{180}\cdot W^2\right.\nonumber\\\nonumber\\
    &&\left.\,+\,\dfrac{(n-12)(n+4)(n^3-244n-120)}{360\,n(n^2-4)}\cdot Q_{4,n}\right\}~,
\end{eqnarray}
\end{widetext}
and finally, by the conformal principle, we can complete the local information on the heat coefficient
\begin{widetext}
\begin{eqnarray}
    (4\pi)^{n/2}\,4\,b_{4,n}[{\cal P}_8]\,=\,&&\dfrac{\Gamma(\frac{n-4}{8})}{\Gamma(\frac{n-4}{2})}\left\{\dfrac{(n-12)(n+4)(n^3-244n-120)}{360\,n(n^2-4)}\cdot Q_{4,n}\,+\,\frac{1}{180}\cdot W^2\nonumber\right.\\\nonumber\\
    &&\left.\,-\,\frac{(n-12)(n-20)(n+4)(n+12)}{360\,n(n-2)}\cdot\left[\Delta J+\frac{n-4}{2}\,J^2\right]\right\}~.
\end{eqnarray}
\end{widetext}
The only available cross check of the above result, to our knowledge, is provided by the 4D trace anomaly 
\begin{equation}
   (a, c-a,\gamma)= (-\frac{172}{45},\frac{1}{60},-\frac{256}{45})~,
\end{equation}
where we can verify, besides the central charges $a$ and $c-a$, our expectation for $\gamma$ from the Cappelli-Coste relation and the {\it improved} Casimir energy, namely $\left.-\frac{k^5}{45}\right|_{k=4}=-\frac{256}{45}$. A careful examination ought to produce a result for generic k, as we will shortly see.

\section{Heat coefficient $b_{4,n}$ : take II}
The key observation, that will allow for a generalization of the previous computations, is that Gilkey's formula does not require a full knowledge of the operator, it is enough to get a handle on $p_2$ and $p_4$:  
\begin{equation}
{\cal P}_{2k}\,=\,(\Delta)^k\,+\,p_{2,ij}\nabla^ i\nabla^j(\Delta)^{k-2}\,+\,(-1)^k\left\{p_{4,i_1...i_{k-4}}\nabla_{i_1}...\nabla_{i_{2k-4}}\right\}\,+...~.
\end{equation}
Moreover, $p_4$ is required only up to terms that upon trace become a total derivative or involving the obstruction tensors. On dimensional grounds, only the Bach tensor would be a candidate, but since it is traceless and divergence-less, it will not show up in $p_4$,

\begin{eqnarray}
    (4\pi)^{n/2}\,k\,B_{4,n}[{\mathcal{ P}}_{2k}]\,=\,\dfrac{\Gamma(\frac{n-4}{2k})}{\Gamma(\frac{n-4}{2})}\cdot\frac{1}{360\,k^2\,n(n^2-4)}& & \int_{\mathfrak{M}_n}dvol_g\,\left\{k^2\,n(n^2-4)\left[2\,Riem^2\,-\,2\,Ric^2\,+\,5\,R^2\right]\right. \nonumber\\
    &+&\,60\,k\,n(n+2)\,R\cdot p_2,_i^{\;i}\nonumber\\
    &-&120\,k\,n(n+2)\,R^{ij}\cdot p_2,_{ij}\,+\,180\,(2k+n-4)(p_2,_i^{\;i})^2\nonumber\\
    &+& \left.360\,(2k+n-4) |p_2|^2\,-\,720\,k\,n(n+2)\,S(p_4)/S(\delta^{k-2})\right\}~.
\end{eqnarray}

We will now extract this relevant information from Juhl's explicit formula.   
The generic expression for $p_2$ can easily be derived from Juhl's paper (see also Michel~\cite{michel2010massedesoperateursgjms}) from terms {\it with strictly more than $2k-4$ derivatives}. Our only challenge is working out the contribution from terms {\it with exactly $2k-4$ derivatives} that will enter the $p_4$ coefficient.

Consider first the formula obtained by Juhl~\footnote{We correct a misprint, a minus sign, in the coefficient of $P$.}
\begin{equation}
\Delta^k\,+\,\sum_{l=0}^{k-2}\Delta^l\nabla^i\left(4(l+1)(k-l-1)P_{ij}-(n-2)Jg_{ij}\right)\nabla^j\Delta^{k-l-2}\,+\,\frac{n-2}{2}\sum_{l=1}^{k-2}\Delta^l\left(J\Delta^{k-l-1}\right)~.
\end{equation}
To obtain $p_2$ it is enough to shift all derivatives to the right, for otherwise only contributions 
to $p_3$ or $p_4$ will be produced,
\begin{eqnarray}
p_{2,ij}\,&=&\,\sum_{l=0}^{k-2}\left(4(l+1)(k-l-1)P_{ij}-(n-2)Jg_{ij}\right)\,+\,\frac{n-2}{2}\sum_{l=1}^{k-2}J\,g_{ij}\nonumber\\\nonumber\\
&=&\,4\binom{k+1}{3}P_{ij}\,-\,\binom{k}{1}\frac{n-2}{2}Jg_{ij}~.
\end{eqnarray}
The above is equivalent to realizing that the $p_2$ coefficient can only come from terms in ${\cal P}_{2k}$ of the form
\begin{equation}
\alpha_1\cdot \mu_2\Delta^{k-1}\,+\,\alpha_2\cdot 4P_{ij}\nabla^i\nabla^j\Delta^{k-2}~.
\end{equation}

\subsubsection{$\alpha_1$:} To determine $\alpha_1$, it is easy to verify by examination of the explicit formula that the only source is ${\cal M}_2^k$, this appears only once with coefficient $n_I=1$ for $I=(1,...,1)$ of length $k$. The power ${\cal M}_2^k=(\Delta+\mu_2)^k$ renders $k\cdot\mu_2\cdot\Delta^{k-1}$. We keep the binomial symbol just for aesthetics, although it is evident that we are picking out one out of the list of k terms.

\subsubsection{$\alpha_2$:} The piece with $\alpha_2$, on the other hand, comes from one ${\cal M}_4$ and ${\cal M}_2^{k-2}$. These terms appear from $I=(1,...,1,2,1,...,1)$ with length $k-1$ and coefficient $n_I=1\cdot...\cdot1\cdot\binom{j}{1}\cdot\binom{k-j}{1}\cdot1\cdot...\cdot1$, where $j=1,..,k-1$ denotes the position of the "2" (i.e., of ${\cal M}_4$ in the composition). Each composition contributes $4P_{ij}\nabla^i\nabla^j\Delta^{k-2}$ with the overall coefficient from the combinatorial sum being
\begin{equation}
    \sum_{j=1}^{k-1}\binom{j}{1}\cdot\binom{k-j}{1}\,=\,\binom{k+1}{3}~.\\
\end{equation}
In all, the $p_2$ term is readily determined by 
\begin{equation}
\binom{k}{1}\cdot\mu_2\Delta^{k-1}\,+\,\binom{k+1}{3}\cdot 4P_{ij}\nabla^i\nabla^j\Delta^{k-2}~.
\end{equation}

This combinatorial exercise is best suited to work out the $p_4$ coefficient but it requires more effort. By examination, is follows that the $p_4$ can only stem from  terms in ${\cal P}_{2k}$ of the form
\begin{equation}
\alpha_3\cdot \mu_2^2\Delta^{k-2}\,+\,\alpha_4\cdot \mu_4\Delta^{k-2}\,+\,\alpha_5\cdot\mu_2\, 4P_{ij}\nabla^i\nabla^j\Delta^{k-3}\,+\,\alpha_6\cdot48P_{i}^mP_{jm}\nabla^i\nabla^j\Delta^{k-3}\,+\,\alpha_7\cdot\,16P_{ij}P_{kl}\nabla^i\nabla^j\nabla^k\nabla^l\Delta^{k-4}~.
\end{equation}

\subsubsection{$\alpha_3$:} This term, analogous to $\alpha_1$, comes from ${\cal M}_2^k$ but this time we choose two out of the list of length k, which results in 
\begin{equation}
    \alpha_3\,=\,\binom{k}{2}~.
\end{equation}

\subsubsection{$\alpha_4$:} This has the same origin, and counting, as $\alpha_2$, namely one  ${\cal M}_4$ and ${\cal M}_2^{k-2}$, just that this time we keep $\mu_4$ from ${\cal M}_4$
\begin{equation}
    \alpha_4\,=\,\alpha_2\,=\binom{k+1}{3}~.
\end{equation}

\subsubsection{$\alpha_5$:} 
This also has the same origin, and counting, as $\alpha_2$, namely one  ${\cal M}_4$ and ${\cal M}_2^{k-2}$, just that this time we pick up one $\mu_2$ out of the $(k-2)$ ${\cal M}_2$
\begin{equation}
    \alpha_5\,=\,\binom{k-2}{1}\cdot\alpha_2\,=\,4\binom{k+1}{4}~.
\end{equation}

\subsubsection{$\alpha_6$:}  This term comes from one ${\cal M}_6$ and ${\cal M}_2^{k-3}$. These terms appear from $I=(1,...,1,3,1,...,1)$ with length $k-2$ and coefficient $n_I=1\cdot...\cdot1\cdot\binom{j+1}{2}\cdot\binom{k-j}{2}\cdot1\cdot...\cdot1$, where $j=1,...,k-2$ denotes the position of the "3" (i.e., of ${\cal M}_6$ in the composition). Each composition contributes $48P_{i}^mP_{jm}\nabla^i\nabla^j\Delta^{k-3}$ with the overall coefficient from the combinatorial sum being
\begin{equation}
    \alpha_6\,=\,\sum_{j=1}^{k-2}\binom{j+1}{2}\cdot\binom{k-j}{2}\,=\,\binom{k+2}{5}~.\\
\end{equation}

\subsubsection{$\alpha_7$:} This last term comes from the composition of two ${\cal M}_4$ and $(k-4)$  ${\cal M}_2$. They appear from \[I=(1,...,1,2,1,...,1,2,1,...,1)\] 
with length $k-2$ and coefficient $n_I=1\cdot...\cdot1\cdot\binom{i}{1}\binom{k-i}{1}\cdot1\cdot...\cdot1\cdot\binom{j+1}{1}\binom{k-j-1}{1}\cdot1\cdot...\cdot1$, where $i=1,...,k-3$ and $j=i+1,...,k-2$ ($j>i$) denote the position of the two "3's" (i.e., of the two ${\cal M}_6$ in the composition). Each composition contributes $16P_{ij}P_{kl}\nabla^i\nabla^j\nabla^k\nabla^l\Delta^{k-4}$ with the overall coefficient from the combinatorial sum being
\begin{equation}
   \alpha_7\,=\,\sum_{i=1}^{k-3}\sum_{j=i+1}^{k-2}\binom{i}{1}\cdot\binom{k-i}{1}\cdot\binom{j+1}{1}\cdot\binom{k-j-1}{1}\,=\,10\binom{k+2}{6}\,-\,\binom{k+1}{5}~.
\end{equation}

In all, the $p_4$ term is finally determined by 
\begin{eqnarray}
&&\qquad\qquad \binom{k}{2}\cdot\mu_2^2\,\Delta^{k-2}\,+\,\binom{k+1}{3}\cdot\mu_4\,\Delta^{k-2}\,+\,\binom{k+1}{4}\cdot16\,\mu_2\,P_{ij}\,\nabla^i\nabla^j\,\Delta^{k-3}\\\nonumber\\\nonumber
&&\,+\,\binom{k+2}{5}\cdot48\,P_{i}^m\,P_{jm}\,\nabla^i\nabla^j\,\Delta^{k-3}\,+\,\left[\binom{k+2}{6}-\dfrac{1}{10}\binom{k+1}{5}\right]\cdot\,160\,P_{ij}\,P_{kl}\,\nabla^i\nabla^j\nabla^k\nabla^l\,\Delta^{k-4}~.
\end{eqnarray}
The corresponding symmetrized trace $S(p_4)/S(\delta^{k-2})$ can be computed with the following recursive relation (cf. lemma 1.3 in Gilkey's paper) 
\begin{equation}
S(\theta\,\delta)\,=\,\frac{n+2m}{1+2m}\,S(\theta)~,
\end{equation}
where $\theta$ has $2m$ indices, until we hit the four or the two explicit indices
of the Schouten tensors. We obtain
\begin{eqnarray}
\frac{S(p_4)}{S(\delta^{k-2})}\,=\,&&\binom{k}{2}\cdot\mu_2^2\,+\,\binom{k+1}{3}\cdot\mu_4\,-\,\binom{k+1}{4}\cdot\frac{16}{n}\,\mu_2\,J\,+\,\binom{k+2}{5}\cdot\frac{48}{n}\,|P|^2\nonumber\\\nonumber\\
&&+\,\left[\binom{k+2}{6}-\dfrac{1}{10}\binom{k+1}{5}\right]\cdot\frac{160}{n(n+2)}\,\left(J^2+2|P|^2\right)~.
\end{eqnarray}
For the remaining necessary contractions, which we use as partial checks for the generalization, we obtain
\begin{widetext}
\begin{eqnarray}
p_2,_i^{\;i}\,&&=\,-\frac{k}{6}(3n^2-6n-4k^2+4)J \\
(p_2,_i^{\;i})^2\,&=& \,\frac{k^2}{36}(3n^2-6n-4k^2+4)^2\,J^2 \nonumber\\
|p_2|^2\,&=&\,\frac{4}{9}k^2(k-1)^2(k+1)^2|P|^2\,+\,\frac{k^2}{12}(n-2)(3n^2-6n-8k^2+8)\,J^2\nonumber\\ 
R\cdot p_2,_i^{\;i}\,&=&-\frac{k}{3}(n-1)(3n^2-6n-4k^2+4)J^2\quad,\quad R^{ij}\cdot p_{2,ij}=\frac{2}{3}(n-2)k(k-1)(k+1)|P|^2 \nonumber\\&-&\frac{1}{3}k(3n^2-9n+2k^2+4)J^2~.\nonumber
\end{eqnarray}
\end{widetext}
In the suitable Q-curvature basis, the integrated heat coefficient reads
\begin{widetext}
\begin{eqnarray}
    (4\pi)^{n/2}\,k\,B_{4,n}[{\mathcal{P}}_{2k}] = \dfrac{\Gamma(\frac{n-4}{2k})}{\Gamma(\frac{n-4}{2})} & & \int_{\mathfrak{M}_n}dvol_g \left\{-\frac{(n-4)(n-2k-4)(n-4k-4)(n+2k-4)(n+4k-4)}{720\,n(n-2)}\cdot J^2 \right. \nonumber\\ &+& \frac{1}{180}\cdot W^2\\
    &-&\left. \dfrac{(n-2k-4)(n+2k-4)(16k^2n-n^3+8k^2-12n-8)}{360\,n(n^2-4)}\cdot Q_{4,n}\right\}~,
\end{eqnarray}
\end{widetext}
and now, by the conformal principle, we complete the local information on the heat coefficient that constitutes the main result of this paper 
\begin{widetext}
\begin{eqnarray}
    (4\pi)^{n/2}\,k\,b_{4,n}[{\cal P}_{2k}]\,=\,\dfrac{\Gamma(\frac{n-4}{2k})}{\Gamma(\frac{n-4}{2})}&&\left\{-\,\dfrac{(n-2k-4)(n+2k-4)(16k^2n-n^3+8k^2-12n-8)}{360\,n(n^2-4)}\cdot Q_{4,n}\,+\,\frac{1}{180}\cdot W^2\right.\nonumber\\\nonumber\\
    &&\left.\,-\frac{(n-2k-4)(n-4k-4)(n+2k-4)(n+4k-4)}{360\,n(n-2)}\cdot\left[\Delta J+\frac{n-4}{2}\,J^2\right]\right\}~.
\end{eqnarray}
\end{widetext}
The available cross-check of the above result, to our knowledge, is again provided by the 4D trace anomaly 
\begin{equation}
   (a\;,\; c-a \;,\;\gamma)= \left(\frac{k^3}{144}-\frac{k^5}{240}\;,\;\frac{k}{180}\;,\;-\frac{k^5}{45}\right)~,
\end{equation}
where we can verify, besides the central charges $a$ and $c-a$, our expectation for $\gamma$ from the Cappelli-Coste relation and the {\it improved} Casimir energy.

\section{Conclusion}
In all, we have succeeded in computing the diagonal $b_{4,n}$ heat-kernel coefficient for the entire family of GJMS operators, completing the program initiated by Branson for the Paneitz operator. We expect that our explicit results for these higher derivative operators, apart from independent mathematical interest, will prove useful in testing recent novel computational methods in the physics literature~\cite{Barvinsky:2024irk,Barvinsky:2024kgt,Barvinsky:2021ijq}.  
We have also been able to verify the rather intricate prediction for the total derivative term in the trace anomaly. In light of the Branson-{\O}rsted conformal principle, things are more transparent though: the total derivative goes hand in hand with its conformal primitive $\left[\Delta J+\frac{n-4}{2}\,J^2\right]$ and any discrepancy in the total derivative coefficient is automatically reflected in the conformal primitive and in the corresponding Polyakov formula. Even if the total derivative vanishes because the manifold under consideration has constant curvature, as in the case of $S^4$ or $S^1_\beta\times S^3$, the partition function is sensitive to its coefficient. 

Interestingly, Barvinsky and Kalugin~\cite{Barvinsky:2024kgt} have recently argued that the divergent part of the multiplicative anomaly should be given by a total derivative term. Our present discussion goes along their line, but is taken with a grain of salt. The total derivative term does not play a role in closed manifolds, but the vanishing coefficient in front of its conformal primitive, as follows from the Branson-{\O}rsted principle for conformally covariant operators or powers thereof, cancels against the pole and results in the finite contribution that matches the finite multiplicative anomaly computed by spectral methods~\cite{Dowker:2010qy,Aros:2023hgi}. 

Let us close by mentioning that a similar prediction for total derivatives and Casimir energy can be established in 6D~\cite{Aros:2023hgi,Aros:2024zxd}. The corresponding computation of the heat kernel coefficient $b_{6,n}$ for higher derivative operators, despite much progress in computer-aided symbolic manipulations, remains a challenge.

\begin{acknowledgments}
The authors are grateful to A. Barvinsky, L. Casarin, and O. Zanusso for valuable correspondence and comments. This work was partially funded through FONDECYT-Chile 1220335. D.E.D. thanks the organizers of the XXI International Congress on Mathematical Physics 2024 in Strasbourg for the kind invitation to contribute an article to a Special Topic in the Journal of Mathematical Physics. 
\end{acknowledgments}


%

\end{document}